\pdfoutput=1

\documentclass[aps,prl,reprint,groupedaddress, floatfix]{revtex4-1}
\usepackage{amsmath}
\usepackage{graphicx}
\usepackage[caption=false]{subfig}
\usepackage{floatrow}
\begin{document}



\title{Learning Unknown Physics of  non-Newtonian Fluids}


\author{Brandon Reyes}
\affiliation{Pacific Northwest National Laboratory, Richland, WA}
\author{Amanda A. Howard}
\affiliation{Pacific Northwest National Laboratory, Richland, WA}
\author{Paris Perdikaris}
\affiliation{University of Pennsylvania, Philadelphia, PA}
\author{Alexandre M. Tartakovsky}
\email[]{amt1998@illinois.edu}
\affiliation{Pacific Northwest National Laboratory, Richland, WA; Department of Civil and Environmental Engineering, University of Illinois Urbana-Champaign, Urbana, IL}


\date{\today}

\begin{abstract}
We extend the physics-informed neural network (PINN) method to learn viscosity models of two non-Newtonian systems (polymer melts and suspensions of particles) using only velocity measurements. The PINN-inferred viscosity models agree with the empirical models for shear rates with large absolute values but deviate for shear rates near zero where the analytical models have an unphysical singularity. Once a viscosity model is learned, we use the PINN method to solve the momentum conservation equation for non-Newtonian fluid flow using only the boundary conditions.
\end{abstract}


\maketitle

\section{Introduction}
In many applications, data is scarce and indirect and the governing physics is not fully known, which limits the utility of standard machine learning (ML) and physics-based methods. For example, in non-Newtonian flow experiments one can easily measure velocity, but not stress or viscosity.  This makes it impossible to use data-driven ML methods to learn stress  as a function of velocity or shear rate. Also, the momentum and mass conservation equations governing non-Newtonian flow are not fully known as one needs to assume a stress-shear-rate relationship (we refer to such relationships as unknown physics) to close the system of these equations. 
It is important to note that standard parameter estimation methods cannot be used for learning unknown physics because the function space is infinite-dimensional.    It is this issue that the physics informed neural network (PINN) method attempts to solve. PINNs use the known underlying structure of physical laws governed by PDEs or ODEs to predict unknown functions or functionals from indirect observations. 
By representing states of the system and hidden physics with neural networks and training using available data subject to the conservation laws, the PINN method can learn unknown physics using sparse and indirect data. 
In the past, the PINN method was used to learn unknown physics in partially unsaturated flow in porous media \cite{Tartakovsky2020WRR} and bioreactors \cite{tipireddy2019comparative}. In this work, we extend the PINN method for estimating the non-Newtonian viscosity based solely on velocity data.

\section{PINN method for non-Newtonian flow models}
Consider a shear flow of a non-Newtonian fluid between two parallel plates satisfying the steady-state momentum conservation equation: 
\begin{equation}
	\label{gov_eq}
	\frac{d}{dy} \left[ \mu(u_y(y)) \frac{d u(y)}{dy} \right] = -C \hspace{15pt} \text{for} \hspace{5pt} y \in \Omega = (0, H),
\end{equation}
where the velocity vector is given by $\mathbf{u} = (u(y), 0, 0)^T$, $u_y \equiv du/dy$ is the shear rate, the viscous stress has the form $\mu(u_y) u_y (y)$, $\mu(u_y)$ is the unknown shear-rate-dependent viscosity, $H$ is the channel width, and $C$ is a force per unit volume.  The fluid velocity $u$ is subject to the no-slip boundary conditions (BCs): 
\begin{equation}
	\label{dirichlet_bc}
	u(0) = 0, \hspace{10pt} u(H) = 0 .
\end{equation}
We consider two cases: no measurements of $\mu$ are available and  some measurements of $\mu$ are present. In both cases we assume that there are $N_u$ measurements of the velocity profile $u(y)$ for $y \in \Omega$: $u^* (y_i)$ for $i = 1, \dots, N_u$.
We approximate the viscosity $\mu (u_y)$ and the velocity $u(y)$ with fully connected feed-forward deep neural networks (DNNs),
$u(y) \approx \hat{u}(y; \theta)$  and  $\mu(u_y(y)) \approx \hat{\mu}(\hat{u}_y(y;\theta); \gamma)$,
where $\theta$ and $\gamma$ are the DNN weights. We train $\hat{u}$ and $\hat{\mu}$ jointly using Eqs. (\ref{gov_eq}) and (\ref{dirichlet_bc}) as constraints. 
This allows us to train $\hat{\mu}$ even without direct measurements of $\mu$.

We note that the DNNs $\hat{u}$ and $\hat{\mu}$ are known non-linear functions of $y$ and $\theta$ and/or $\gamma$. Therefore,  we can analytically compute the DNN derivatives with respect to $y$ and the weights. The former are needed to impose the physical constraints given by Eq. (\ref{gov_eq}), while the latter are required to update the values for the weights in the process known as backpropagation \cite{hirose1991back}. Here, we use automatic differentiation \cite{baydin2017automatic} to compute the derivatives.

Eq. (\ref{gov_eq}) is enforced in the DNN training by forming an additional ``auxiliary" DNN: 
\begin{align}
	\label{gov_NN}
	\hat f(y; \theta, \gamma) = \frac{d}{dy} \left[ \hat{\mu}(\hat{u}_y(y;\theta); \gamma) \frac{d \hat{u} (y; \theta)  }{dy} \right].
\end{align}
We train the DNNs simultaneously by minimizing the loss function
\begin{align}
	\label{loss_mu}
	L(\theta, \gamma) &= \frac{ \omega_1}{N_u} \sum_{i=1}^{N_u} \left[  \hat{u}(y_i; \theta) - u^* (y_i)  \right]^2  \\
	&+ \frac{ \omega_2 }{2}\left[\hat{u}(y=0; \theta)^2 + \hat{u}(y=H; \theta)^2  \right] \nonumber \\ \nonumber
	&+\frac{ \omega_3 }{N_u} \sum_{i=1}^{N_u} \left[ \hat f(y_i; \theta, \gamma) + C \right]^2 \\ \nonumber
	&+ \frac{ \omega_4}{N_\mu} \sum_{i=1}^{N_\mu} \left[  \hat{\mu}({u_y}_i; \gamma) - \mu^* ({u_y}_i)  \right]^2. \nonumber
\end{align}
In $L(\theta, \gamma)$, the first term forces $\hat{u} (y; \theta)$ to match the velocity measurements, the second term forces $\hat{u} (y; \theta)$ to match the Dirichlet BCs, and the third term forces $\hat{u}$ and $ \hat{\mu}$ to satisfy Eq. (\ref{gov_eq}). The last term is present ($\omega_4 \neq 0$) if measurements of $\mu$ (i.e., $\mu^* ({u_y}_i)$ for $i=1,...,N_\mu$) are available and forces $\hat{\mu}$ to match these measurements. The weights $\{\omega_i\}_{i = 1, \ldots, 4}$ reflect the fidelity level of the data and physics models. For example, $u$ measurements are more accurate than viscosity measurements in general, so $\omega_1 \geq \omega_4$. We note that Eq. (\ref{gov_eq}) is an approximation of the momentum conservation equation because it involves assumptions about the general form of the viscous stress, therefore, $\omega_3 \leq \omega_1$. For some flows the no-slip BCs assumption might not be very accurate, which would affect the relative value of $\omega_2$.  The relative values of $\omega_i$ can also affect the convergence rate of iterative solutions of the minimization problem  
$
(\theta, \gamma) = \text{arg} \min_{\theta, \gamma} L(\theta, \gamma) 
$ 
\cite{Wang2020,wang2020and}. 

To solve this minimization problem, we set the initial values of $\theta$ and $\gamma$ using the Xavier's normal initialization scheme \cite{Xavier_scheme}. Next, we run the Adam optimizer \cite{kingma2014adam} for a set number of steps. Finally, we run the quasi-Newton L-BFGS-B optimizer \cite{L_BFGS-B} until the desired convergence and tolerance are achieved. We find that  for the considered here problems, this combination of the optimizers increases the convergence rate and reduces the computational cost as compared to using either optimizer alone. We use DNNs with two hidden layers with sixty nodes each and a learning rate of 0.001 for the Adam optimizer. The error $\| e_{\hat{u}} \|_2 =  \| \hat{u}(y; \theta) - u^* (y) \|_2/ \|  u^* (y) \|_2$ estimates the accuracy of the DNN approximations of $u$ relative to the $u$ measurements and the error $\| f \|_\infty = \max_{1 \leq i \leq N_u} |f(y_i; \theta, \gamma) + C |$ is a measure of how well the DNN approximations of $u$ and $\mu$ satisfy Eq. (\ref{gov_eq}). 

We refer to the PINN method that is used to evaluate the unknown viscosity function given the measurements $u$ (or $u$ and $\mu$) as the inverse PINN.  Once $\hat{\mu}$ is trained, the PINN method can also be used to solve the momentum conservation equation without observations of $u$ ({and/or $\mu$}) if the shear rate does not exceed the maximum shear rate in the experiment used to train $\hat{\mu}$. 
To train $\hat{u}$ as an approximate solution of Eq. (\ref{gov_eq}) we use the loss function Eq. (\ref{loss_mu})  with $\omega_1 = \omega_4 = 0$ and $\omega_2 = \omega_3 = 1$. We refer to this application of PINNs as the forward PINN method. 


\section{Validation of the inverse and forward PINN methods}

We first validate the ability of the inverse PINN method to learn the unknown shear-dependent viscosity using velocity data generated with the Ostwald-de Waele power-law effective viscosity model \cite{bird2006}, 
	$\mu_{pl} \left( u_y(y)\right) = K \left| u_y(y) \right| ^{n-1}$,
where $K$ is the power-law consistency coefficient and $n$ is the power-law index. This model in combination with Eqs. (\ref{gov_eq}) and (\ref{dirichlet_bc}) allows for an analytical solution for $u(y)$ and $du(y)/dy$ \cite{Hinch}. 

\begin{figure}
	{\includegraphics[width=0.32\linewidth]{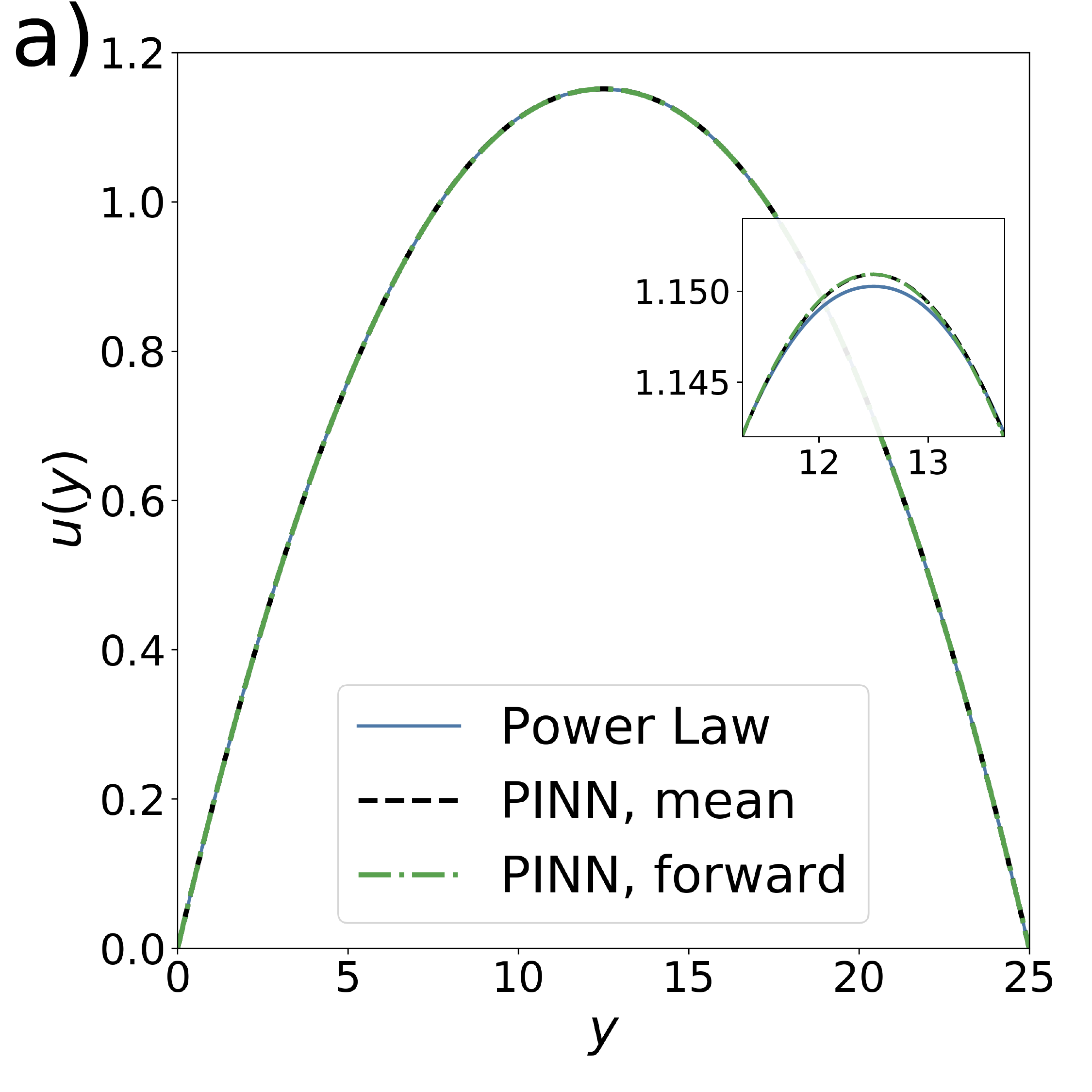}}
	{\includegraphics[width=0.32\linewidth]{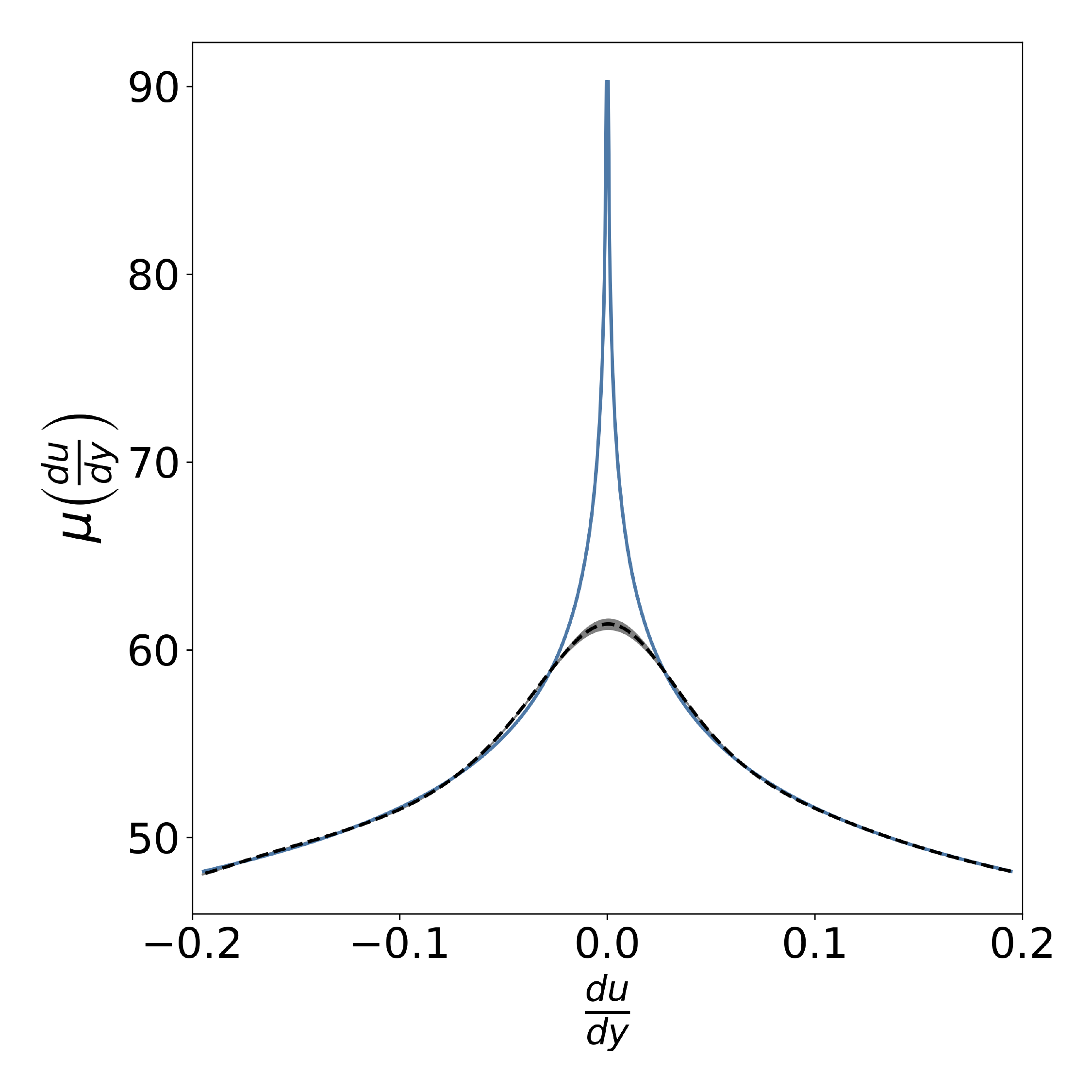}}
	{\includegraphics[width=0.32\linewidth]{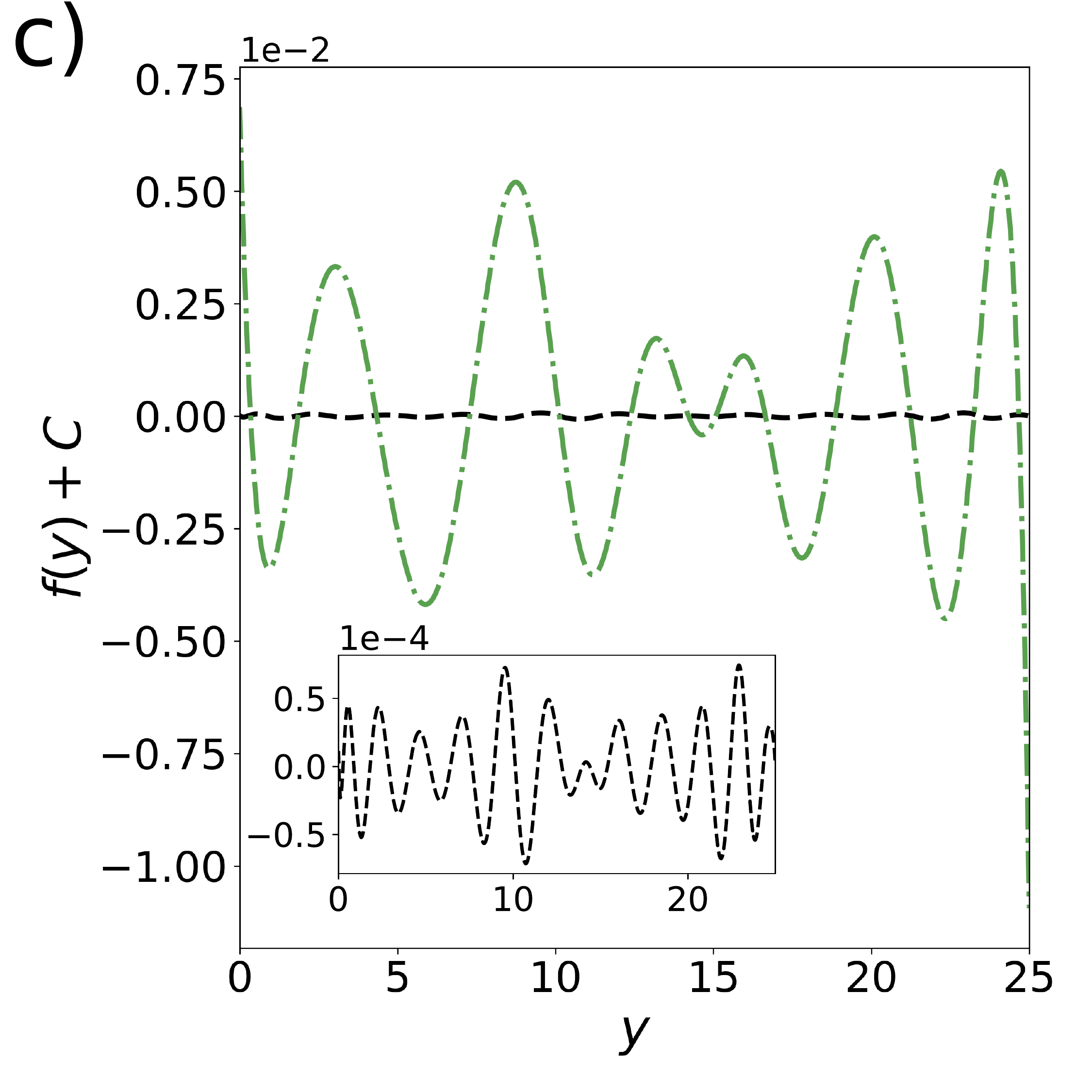}}
 \caption{Inverse and forward PINN solutions for the synthetic data generated from the analytical solution for a power-law fluid with  $n = 0.898$. 
 (a) The average velocity profile from 100 runs for the PINN results, and one run for the PINN forward model. (b) The average $\frac{du}{dy}$ of the 100 runs versus the average $\mu$. The variance is given by the grey area. (c) Average error in satisfying the ODE.  }
\label{fig:analytic}
\end{figure}

We generate two data sets by selecting $N_u =501$ uniformly distributed measurements of $u$ from the analytical solution for $u$ using both $n = 0.898$ (shear-thinning fluid) and $n = 1.2$ (shear-thickening fluid) with $C = 0.75$, $H = 25$, and $K = 40.788$. For both values of $n$ we  train the $\hat{u}(y;\theta)$ and $\hat{\mu}(\hat{u}_y(y;\theta);\gamma)$ DNNs by minimizing the loss function Eq. (\ref{loss_mu}) with $\omega_1 = \omega_2 =\omega_3 =1$ and $\omega_4=0$. We note that the minimization problem is not convex and its solution $(\theta,\gamma)$ can depend on the initial values of  $\theta$ and $\gamma$. To demonstrate how different initial values for the weights affect the PINN solution, we solve the minimization problem with 100 different initializations of $\theta$ and $\gamma$ and then average the resulting DNNs $\hat{u}(y;\theta)$ and $\hat{\mu}(\hat{u}_y(y;\theta);\gamma)$ to obtain the solutions for $u(y)$ and $\mu(u_y)$, respectively. For $n = 0.898$ the average solutions are compared with the analytical solutions in Figs. \ref{fig:analytic}a and  \ref{fig:analytic}b. The average DNN $\hat{u}(y)$ solution agrees well with the analytical $u(y)$ solution. The average DNN $\hat{\mu}(\hat{u}_y)$ solution agrees very well the constitutive model  for large shear rates. For small shear rates, the DNN solution deviates from the analytical solution and for zero shear rate has a finite value while the analytical solution has a nonphysical singularity. 
 Fig. \ref{fig:analytic}b also shows that the standard deviation in the learned $\mu(u_y)$ is largest at $u_y = 0$ and is several orders of magnitude smaller than the mean value of $\mu$ at $u_y = 0$, indicating that the uncertainty of the PINN method due to DNN initialization is relatively small.  
Fig. \ref{fig:analytic}c depicts the residual $\hat{f}(y;\theta,\gamma) + C$ of Eq. (\ref{gov_eq}) as a function of $y$. The small values of the residual show that the DNNs $\hat{u}$ and $\hat{\mu}$ approximately satisfy Eq. (\ref{gov_eq}).    
The $\| e_{\hat{u}} \|_2$ and $\| f \|_\infty$ errors for both values of $n$ are given in Table \ref{tab:beads_error_analytic}. Small errors demonstrate that the inverse PINN method is equally accurate for both shear-thinning and shear-thickening fluids. 

\begin{table}
 \caption{Mean errors computed from 100 inverse PINN solutions for the synthetic data generated from the analytical solution for a power-law fluid with  $n = 0.898$ and 1.2.} \label{tab:beads_error_analytic}
 \begin{ruledtabular}
	\begin{tabular}{l l l}
		$n$ & $\| e_{\hat{u}}\|_2$   & $\|f\|_\infty$  \\ \hline
		0.898 & 2.345$\times10^{-4}$ & 7.45$\times10^{-5}$ \\ 
		1.2 & 2.135$\times10^{-4}$ & 5.061$\times10^{-4}$ \\ 
	\end{tabular}
	\end{ruledtabular}
\end{table}

Next, we validate the ability of the forward PINN method to solve Eq. (\ref{gov_eq}). 
We fix the weights of the DNN $\hat{\mu}$ obtained from the inverse PINN with $n=0.898$ and train the $\hat{u}(y,\theta)$ DNN by minimizing the loss function Eq. (\ref{loss_mu})  with $\omega_1 = \omega_4 = 0$, $\omega_2 = \omega_3 = 1$.  Fig. \ref{fig:analytic}a shows that the trained $\hat{u}(y,\theta)$ closely agrees with the analytical solution for the power-law fluid with $n=0.898$. Fig. \ref{fig:analytic}c  shows the maximum residual corresponding to this DNN is two orders of magnitude smaller than $C$, indicating that $\hat{u}$ approximately satisfies Eq. (\ref{gov_eq}).
The good agreement between $\hat{u}$ and the reference solution for $u$ and small residuals confirm the accuracy of the forward PINN method for solving non-linear differential equations with constitutive relationships given by a DNN with known weights.

\section{Monodisperse polymer melts} 
\label{sec:polymers}

We consider a synthetic Dissipative Particle Dynamic (DPD) fluid consisting of chains of $N$ equal-size beads connected by springs to model polymer melts. Two-dimensional DPD simulations of such fluids between two parallel plates with chains made of $N=2$, 5, and 25 beads are presented in  \cite{RPF_LEC_paper}. In \cite{RPF_LEC_paper}, the DPD results were used to compute  $\mu_{DPD}(du(y)/dy)$ using the Irving-Kirkwood relationship \cite{irving1950statistical}. 

 We use the velocity data from \cite{RPF_LEC_paper} and the inverse PINN method  with $\omega_1 = \omega_2= \omega_3 = 1$ and $\omega_4 = 0$ in Eq. (\ref{loss_mu}) to estimate $\mu(u_y)$. To match \cite{RPF_LEC_paper} , $C = 0.75$ and $H=25$. The relative velocity error and the maximum residual error are given in  Table \ref{tab:beads_error}. For all considered $N$ the relative error in $u$ is less than 0.1\% and the maximum residual error is 3 orders of magnitude smaller than the driving force $C$, indicating that the DNN $\hat{u}$ accurately approximates data and the DNNs $\hat{u}$ and $\hat{\mu}$ satisfy the governing equations. 

\begin{table}
 \caption{Errors for $N = 2$, $5$, and $25$ beads.} \label{tab:beads_error}
 \begin{ruledtabular}
	\begin{tabular}{l l l}
		$N$ & $\| e_{\hat{u}}\|_2$  & $\|f\|_\infty$  \\ \hline
		2 & 2.568$\times10^{-4}$ & 1.544$\times10^{-4}$  \\ 
		5 & 1.921$\times10^{-4}$ & 1.773$\times10^{-4}$ \\
		25 & 4.396$\times10^{-4}$ & 1.201$\times10^{-4}$ 
	\end{tabular}
	\end{ruledtabular}
\end{table}

\begin{figure}
\centering
	{\includegraphics[width=0.32\linewidth]{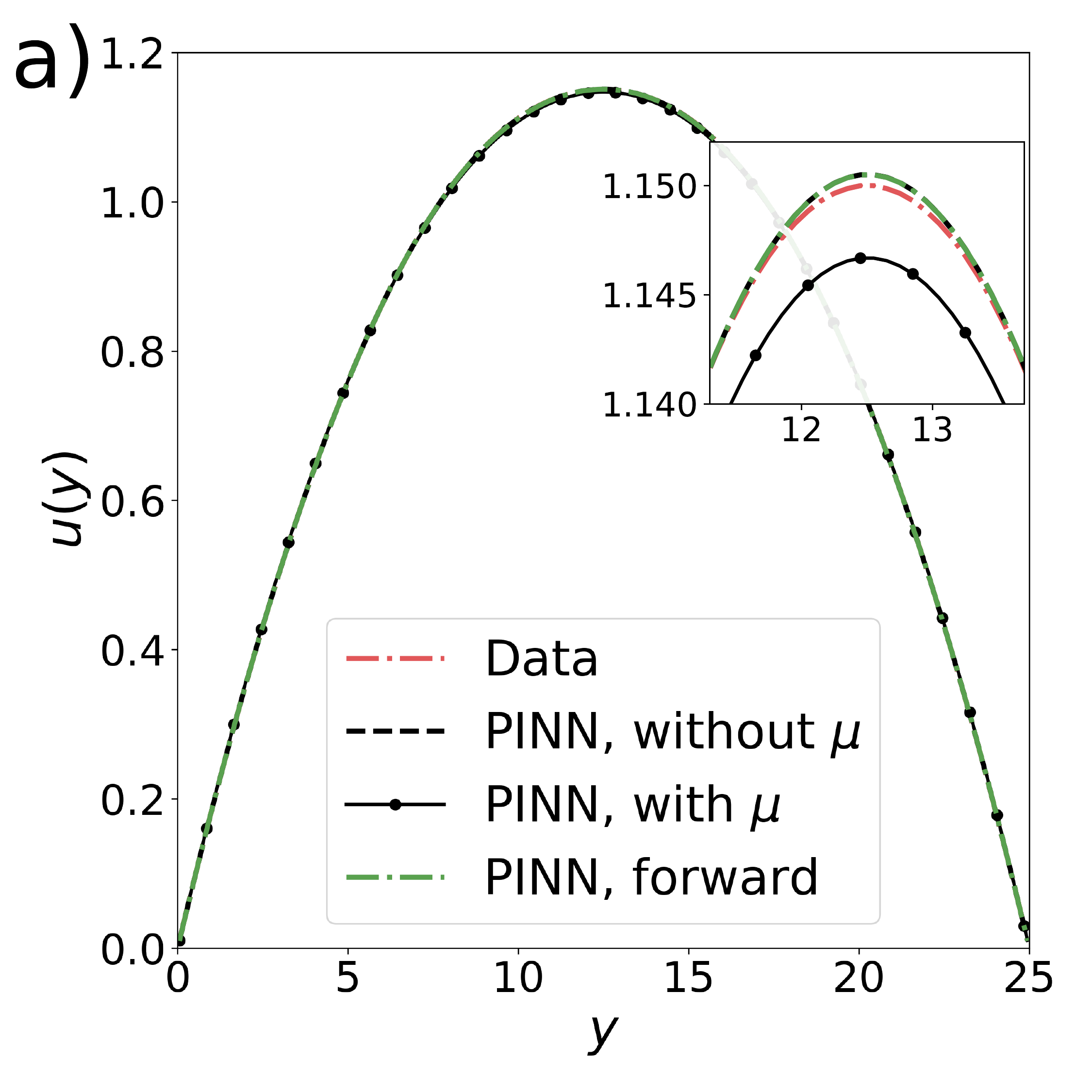}}
	{\includegraphics[width=0.32\linewidth]{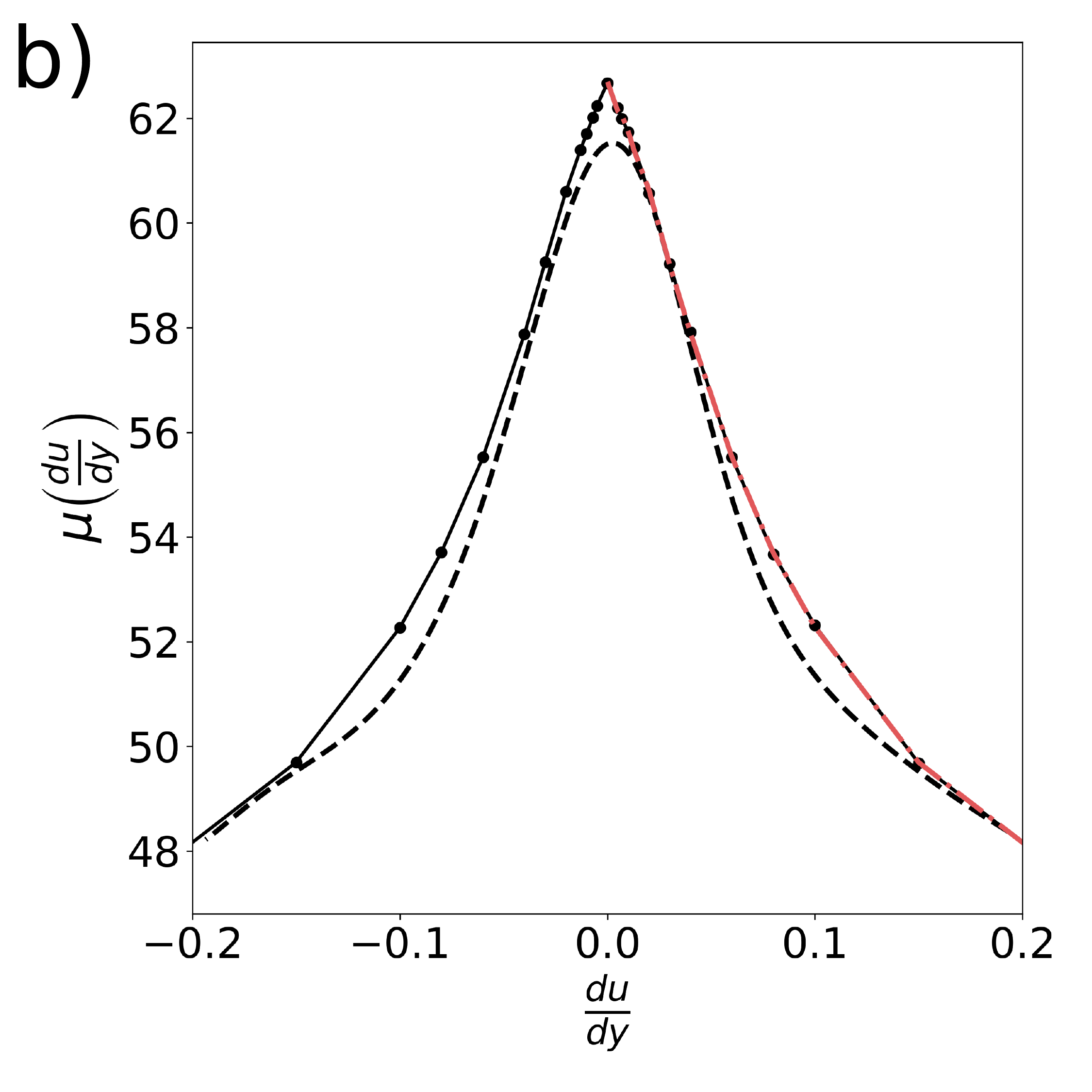}}
	{\includegraphics[width=0.32\linewidth]{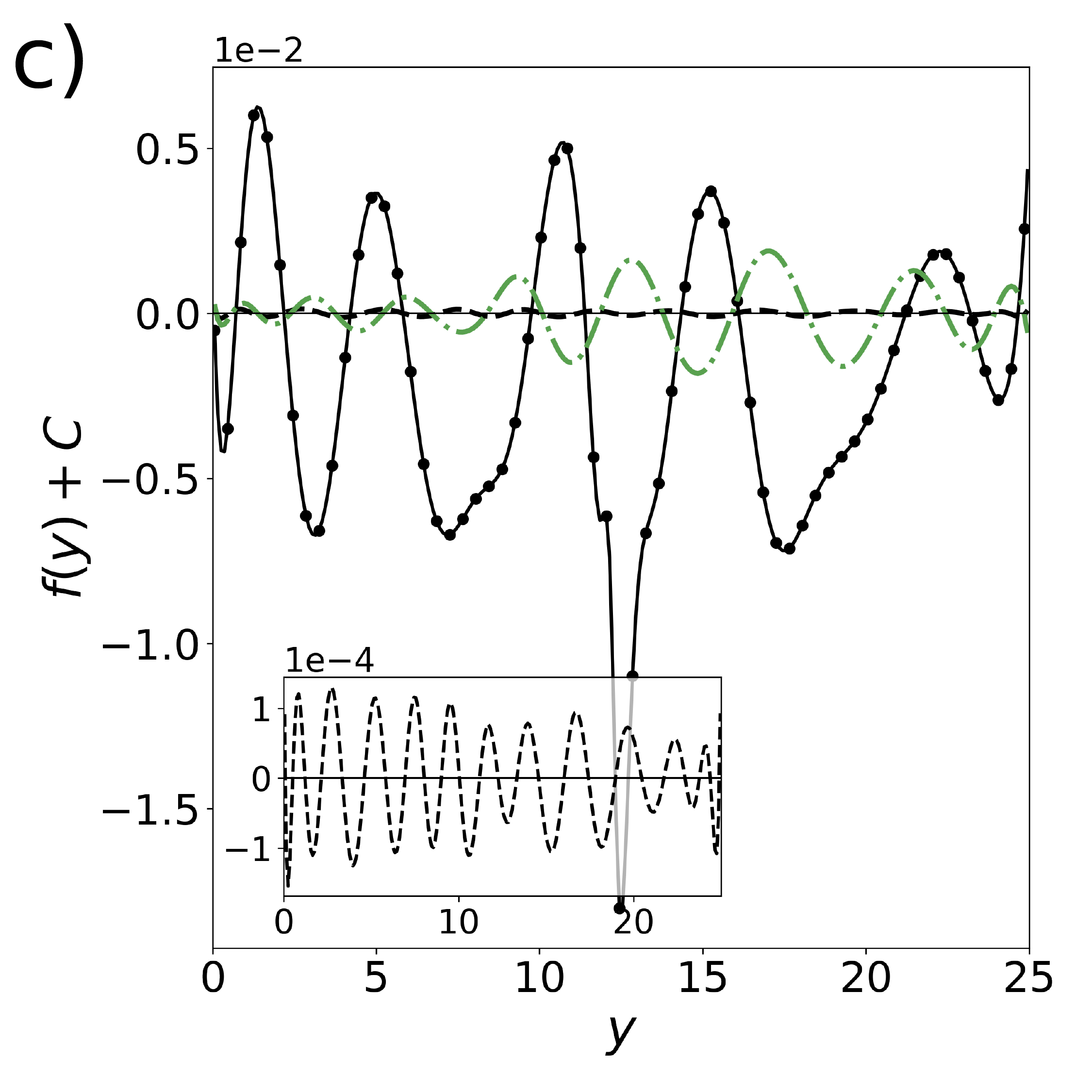}}
 \caption{Inverse and forward model results for $N = 2$. (a) Resulting velocity profiles. (b) Resulting viscosity profile. (c) Error in satisfying the ODE. In the power-law solution we take $n=0.90$ and $K=40.79$. }  \label{fig:N2}
\end{figure}

Figs. \ref{fig:N2}a and \ref{fig:N2}b compare the velocity profiles and viscosities estimated from the DPD simulation, $\mu_{DPD}$ and from the PINN method for $N=2$. The DNN velocity profile $\hat{u}(y;\theta)$ closely matches the DPD velocity profile $u_{DPD}(y)$. The agreement between $\hat{\mu}(u_y;\theta,\gamma)$ and $\mu_{DPD}(u_y)$  is good but less accurate than the agreement for the velocities. To test whether $u_{DPD}(y)$ and $\mu_{DPD}(u_y)$ satisfy Eq. (\ref{gov_eq}), we train the $\hat{u}(y;\theta)$ and  $\hat{\mu}(u_y;\theta,\gamma)$ DNNs conditioned on both $u$ and $\mu$ DPD measurements.
Figs. \ref{fig:N2}a and \ref{fig:N2}b show that  conditioning of the DNNs on the DPD measurements of $u$ and the estimates of $\mu$ produces DNNs that match well both $u$ and $\mu$
data. However, conditioning on the DPD $\mu$ estimates also results in the residual errors that are two orders of magnitude larger than the residual errors in the case where no $\mu$ estimates are used to train the DNNs, as shown in Fig. \ref{fig:N2}c. 

\begin{figure}
\centering
	{\includegraphics[width=0.32\columnwidth]{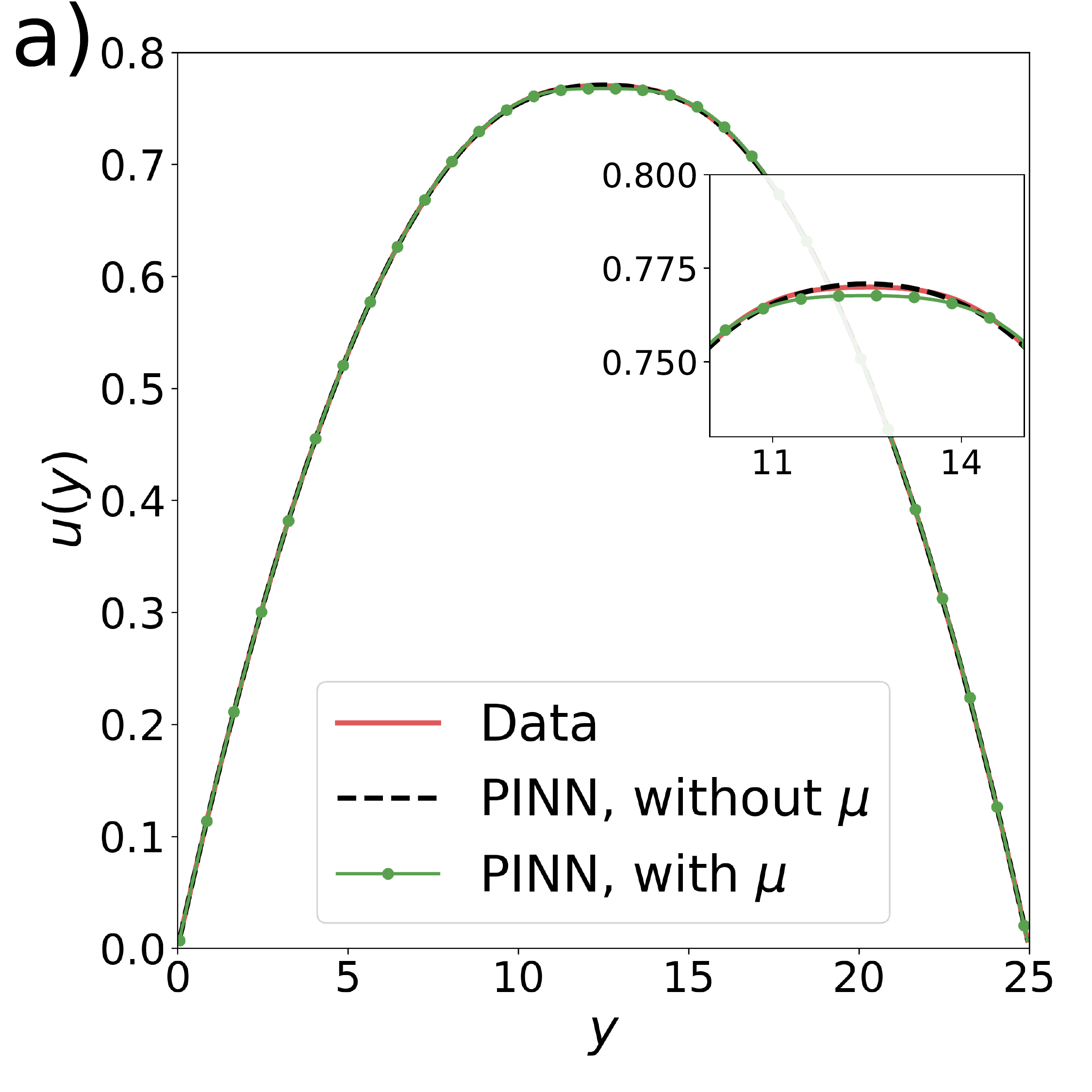}}
	{\includegraphics[width=0.32\columnwidth]{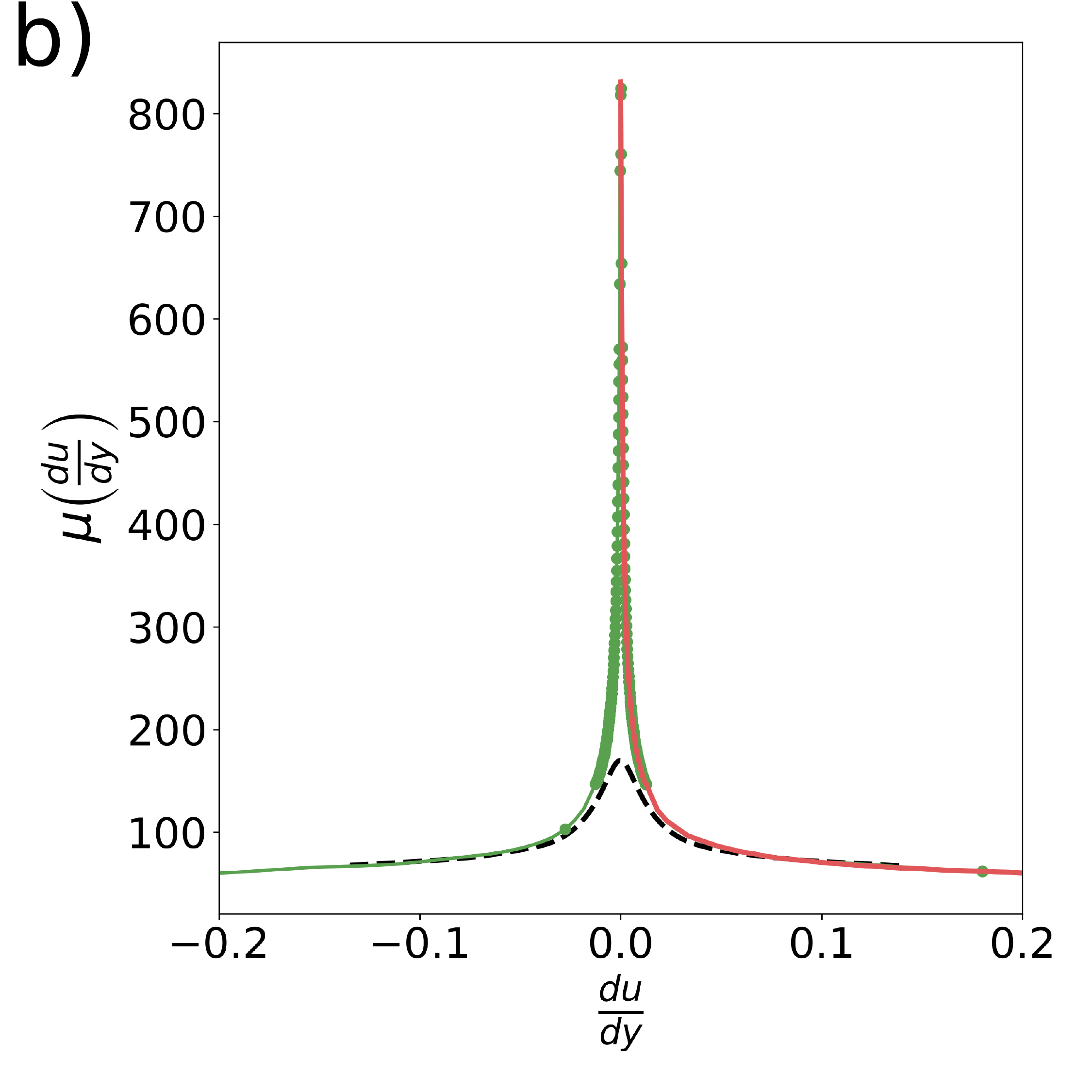}}
	{\includegraphics[width=0.32\columnwidth]{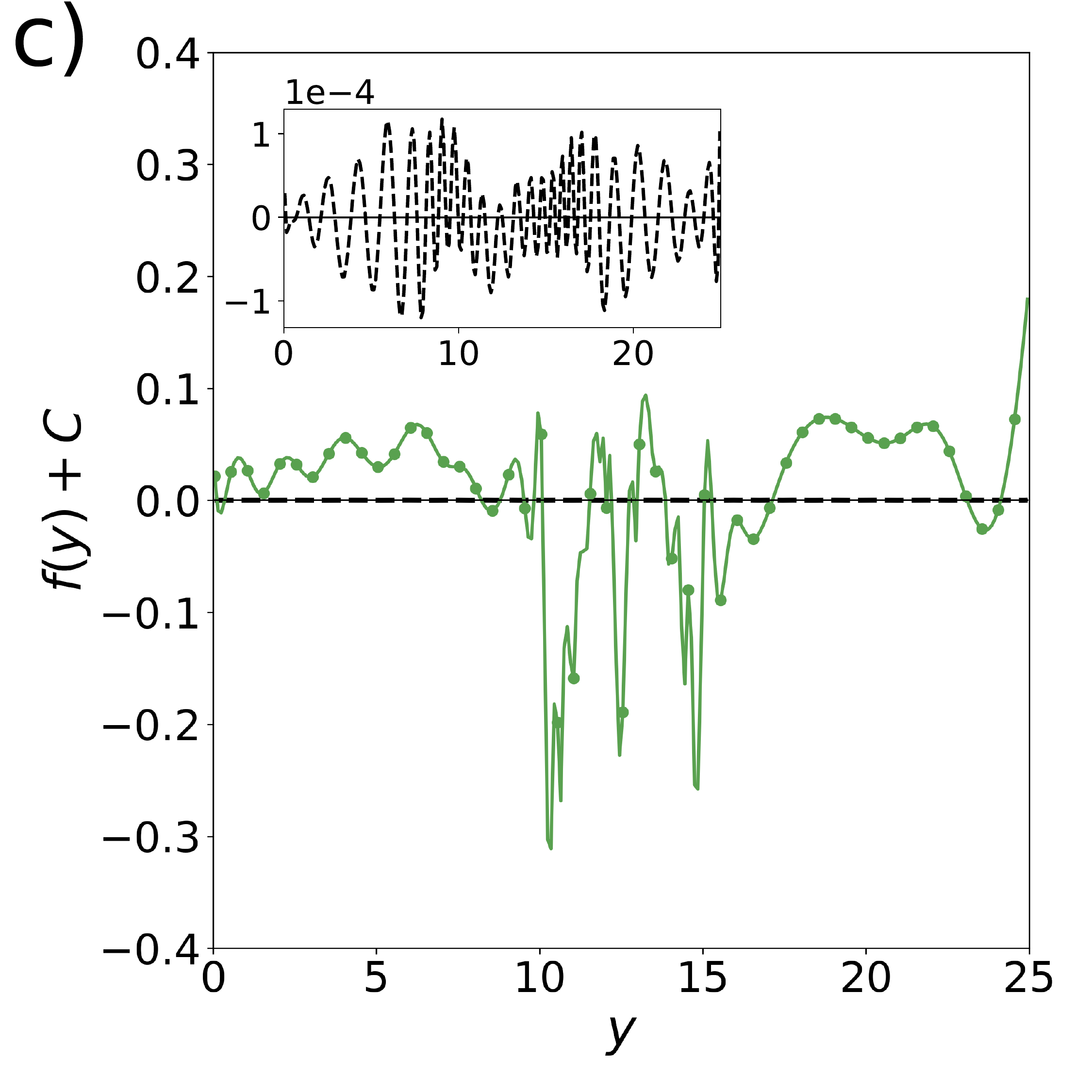}}
 \caption{ PINN results for $N = 25$. (a) Resulting velocity profiles. (b) Resulting viscosity profile. (c) Error in satisfying the ODE. In the power-law solution we take $n=0.79$ and $K=43.78$.}  
   \label{fig:N25}
\end{figure}

Next, we use the PINN method to evaluate the viscosity of the polymer melt with 25-bead chains. As for the melt with $N=2$, we  first train the $\hat{u}$ and $\hat{\mu}$ DNNs using only $u_{DPD}(y)$ measurements.  Fig. \ref{fig:N25} shows that the $\hat{u}$ DNN agrees well with the $u_{DPD}(y)$ measurements and the resulting residual point errors are nearly zero (more than four orders of magnitude smaller than $C$). We also see that $\hat{\mu}$ significantly deviates from the $\mu_{DPD}(u_y)$ values estimated from the DPD simulations near a shear rate of zero. Then, we train the $\hat{u}$ and $\hat{\mu}$ DNNs using both $u_{DPD}(y)$ and $\mu_{DPD}(y)$ data. Fig. \ref{fig:N25}  demonstrates that the resulting DNNs fit the  $u_{DPD}(y)$ and $\mu_{DPD}(y)$ data well, but the corresponding residual is very large (on the order of $C$). We obtain similar results for the polymer melt with $N=5$. 

Finally, we demonstrate that once $\hat{\mu}(u_y;\theta,\gamma)$ is trained, the forward PINN method can be used to solve Eq. (\ref{gov_eq}) subject to the BC Eq. (\ref{dirichlet_bc}). We use the weights $\gamma$ in the $\hat{\mu}(u_y;\theta,\gamma)$ DNN obtained above from the inverse PINN and train the forward solution, $\hat{u}_f(y;\theta)$,  DNN by minimizing the loss function Eq. (\ref{loss_mu})  with $\omega_1 = \omega_4 = 0$ and $\omega_2 = \omega_3 = 1$ for $C = 0.75$. For $N=2$, Fig. \ref{fig:N2}a shows that the $\hat{u}_f(y;\theta)$ DNN matches the experimental data corresponding to  $C = 0.75$ well. In addition to this, Fig. \ref{fig:N2}c demonstrates that the residual of the governing equation is two orders of  magnitude smaller than $C$ confirming that $\hat{u}_f(y;\theta)$ approximately solves Eq. (\ref{gov_eq}) subject to Eq. (\ref{dirichlet_bc}).

These results lead to the conclusion that the inverse PINN is capable of estimating the effective viscosity function $\mu(u_y)$, which can be used for solving the momentum conservation equation (\ref{gov_eq}). The predicted viscosity deviates from the viscosity obtained from the DPD simulations for small shear rates with the discrepancy increasing with the number of beads $N$. Our results show that velocity and viscosity data provided in  \cite{RPF_LEC_paper} cannot accurately be described by Eq. (\ref{gov_eq}).

\section{Dense suspensions of spherical particles}

In this section, we employ the inverse PINN method to learn the shear-rate-dependent viscosity of  densely packed spherical particles suspended in a Newtonian fluid using the velocity measurements presented in  \cite{Howard2018}. 
The considered data are obtained from the numerical simulations of suspension flows in a channel using the Force Coupling Method (FCM) \cite{Yeo2010, Yeo2011, Howard2018}. In the considered suspensions,  the average particle volume fraction $\phi_a = \frac{\frac{4}{3}\pi a^3 N}{V}$ ranges from  0.2 to 0.4, where $a$ is the particle radius, $N$ is the number of particles, and $V$ is the volume of the domain. 
In the FCM simulations, the particle radius was set to $a = 1$, the channel length to $L_x = 80$, the height to $H = 40$, and the width to $L_z = 30$. The channel walls were located at $y=0$ and $40$, constant Dirichlet BCs for pressure were prescribed at the $x=0$ and $80$ boundaries with the pressure drop over the length of the channel $\Delta P/L_x=0.029$, and periodic conditions were used in the $z$ direction. At the continuum level, the considered suspension behaves as a non-Newtonian fluid and can be described by Eq.  (\ref{gov_eq}) with $C = \Delta P/L_x$.

\begin{figure}
\includegraphics[width=0.3\linewidth]{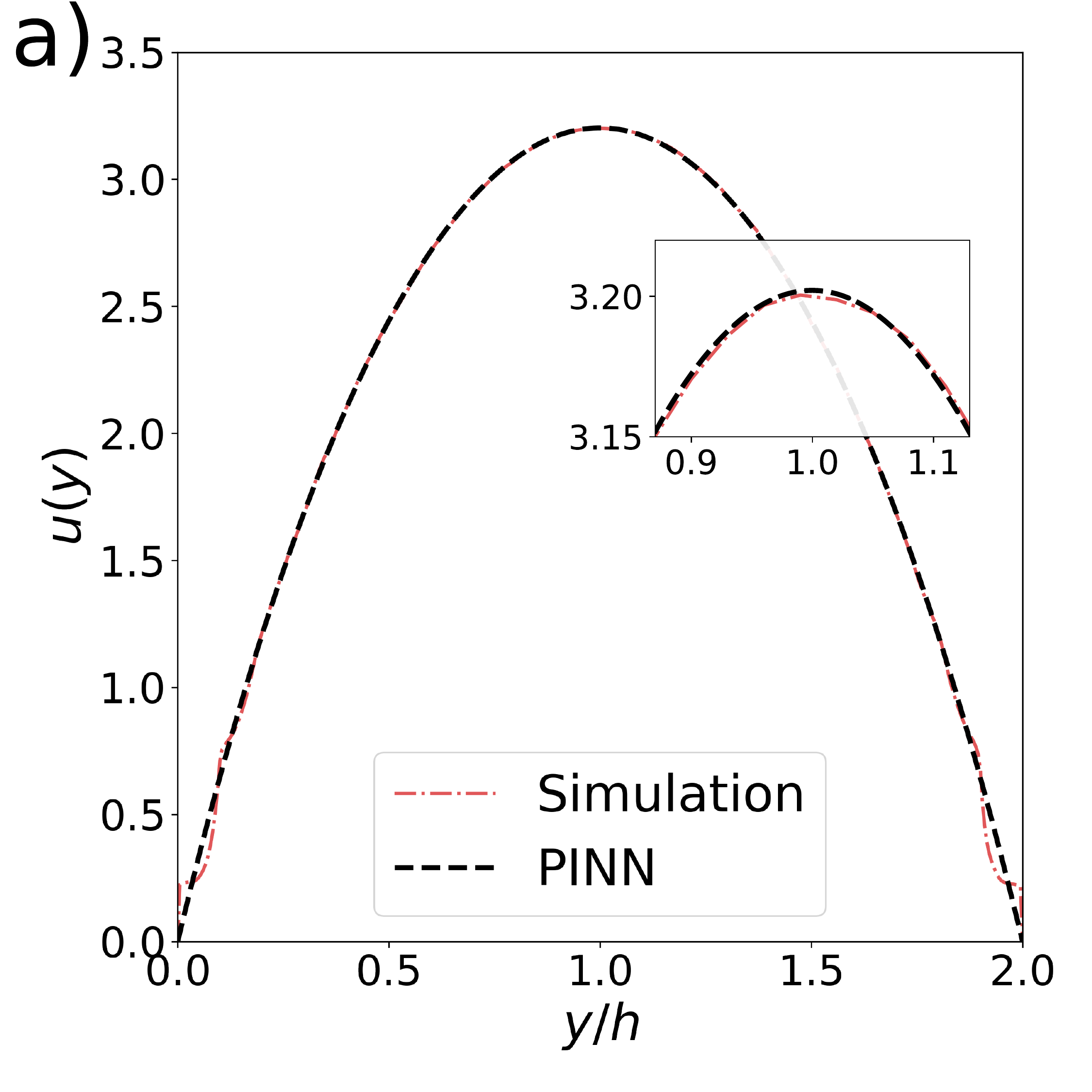}
\includegraphics[width=0.3\linewidth]{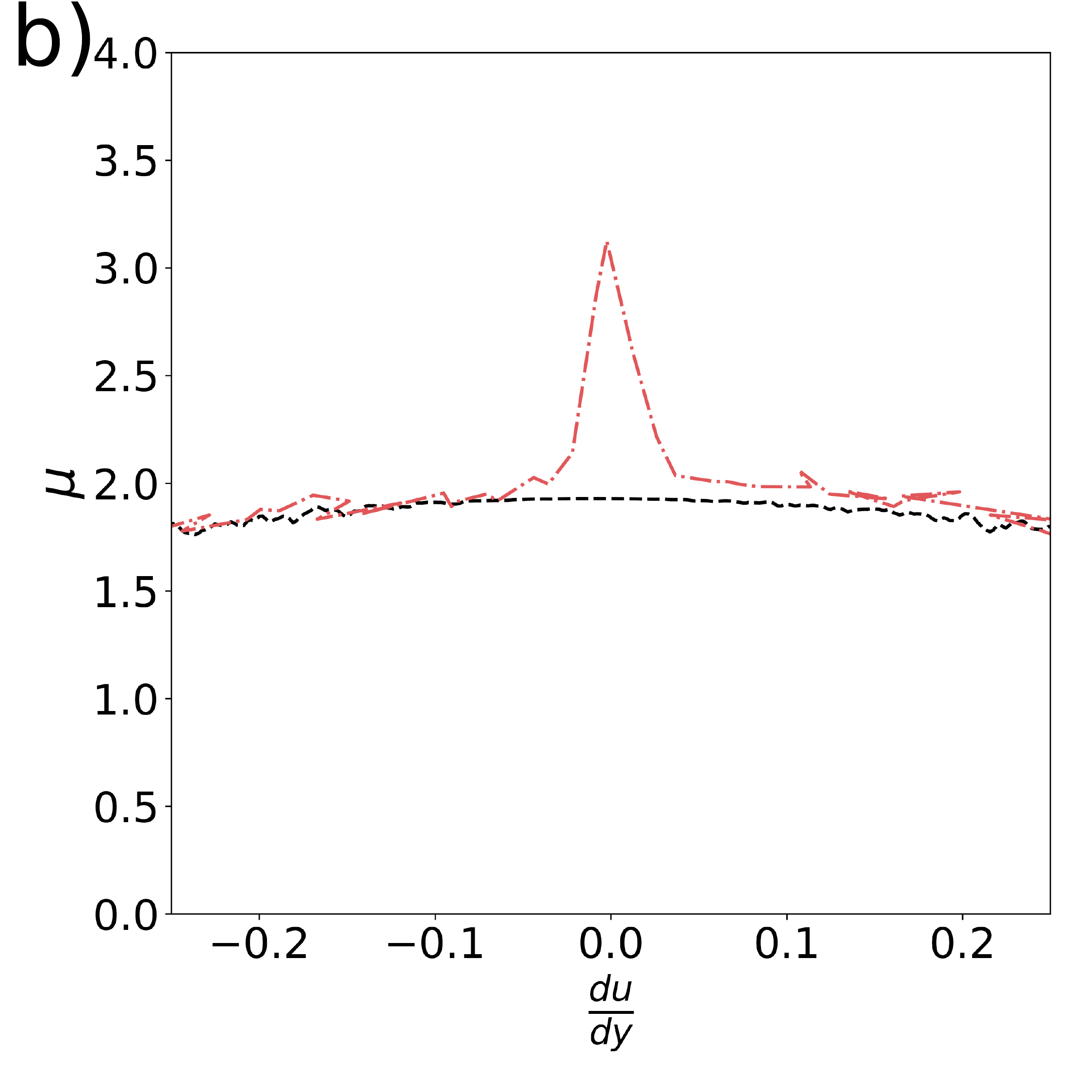}
\includegraphics[width=0.3\linewidth]{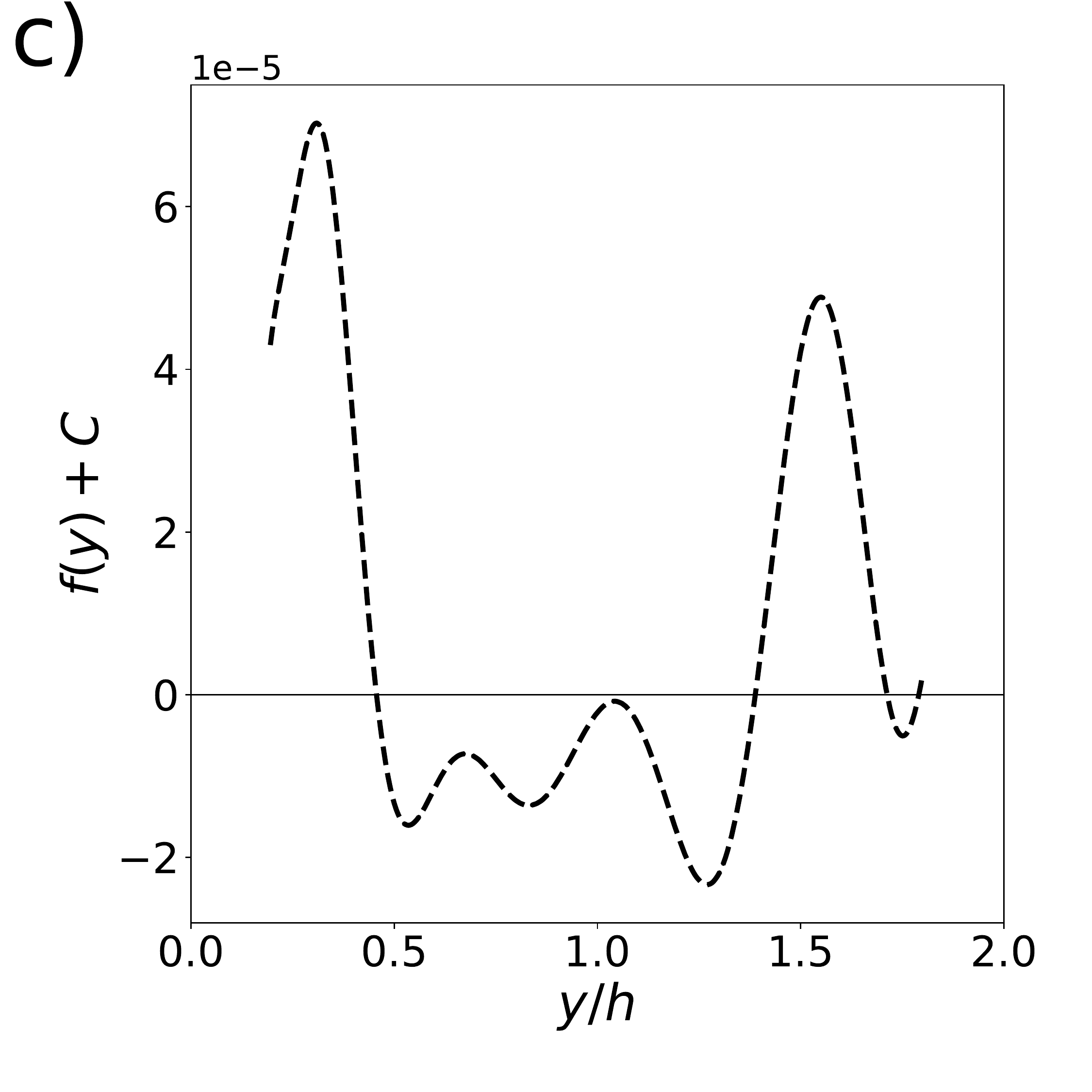}
\\
\includegraphics[width=0.3\linewidth]{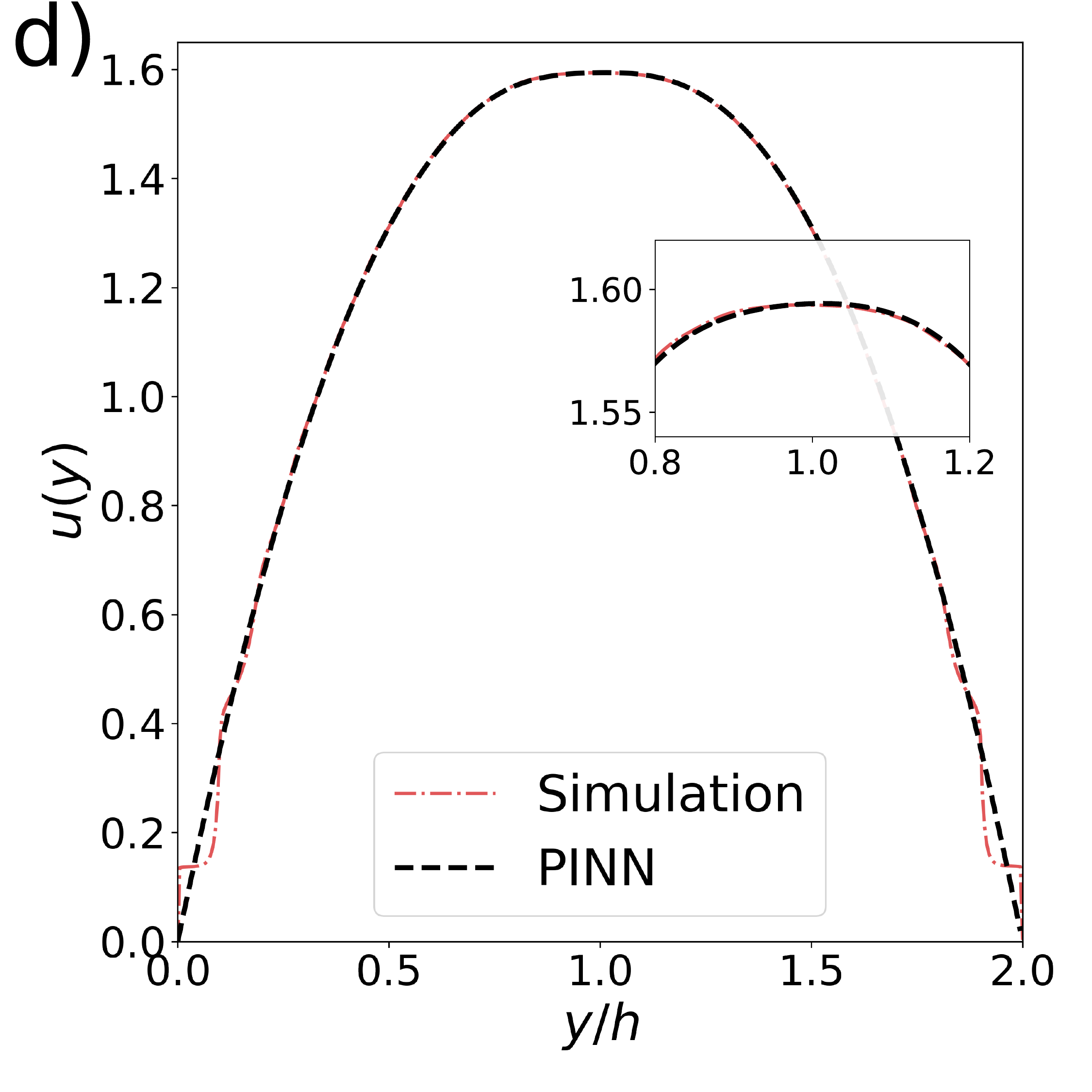}
\includegraphics[width=0.3\linewidth]{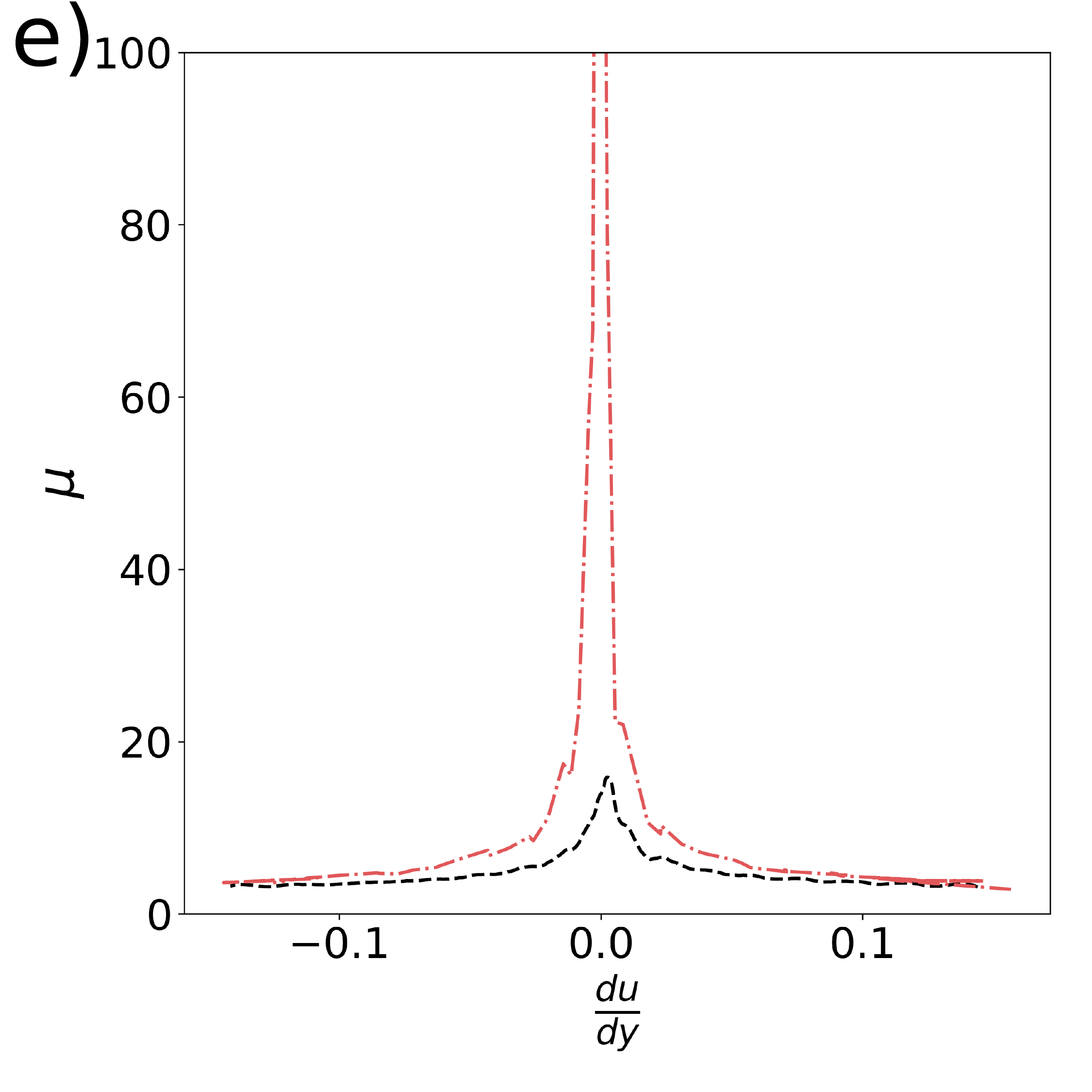}
\includegraphics[width=0.3\linewidth]{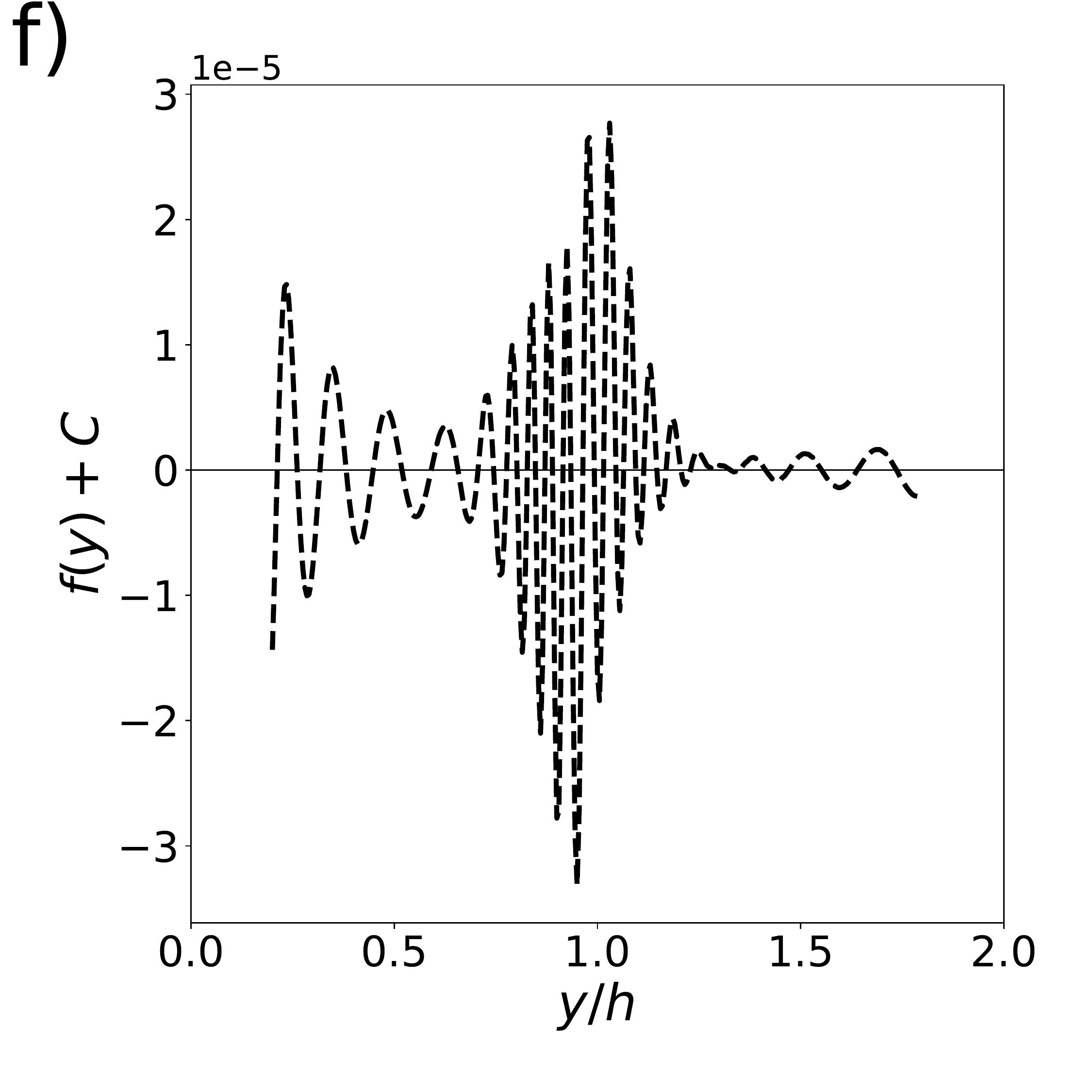}
 \caption{Inverse PINN results for suspensions with average volume fraction $\phi_a=0.2$ (a-c) and $0.4$ (d-f). The velocity profiles, viscosity profiles, and relative errors in satisfying the ODE are shown.  The inverse PINN model is trained with three hidden layers with one hundred nodes each.  \label{fig:suspension}}
\end{figure}

\begin{figure}
{\includegraphics[width=0.45\linewidth]{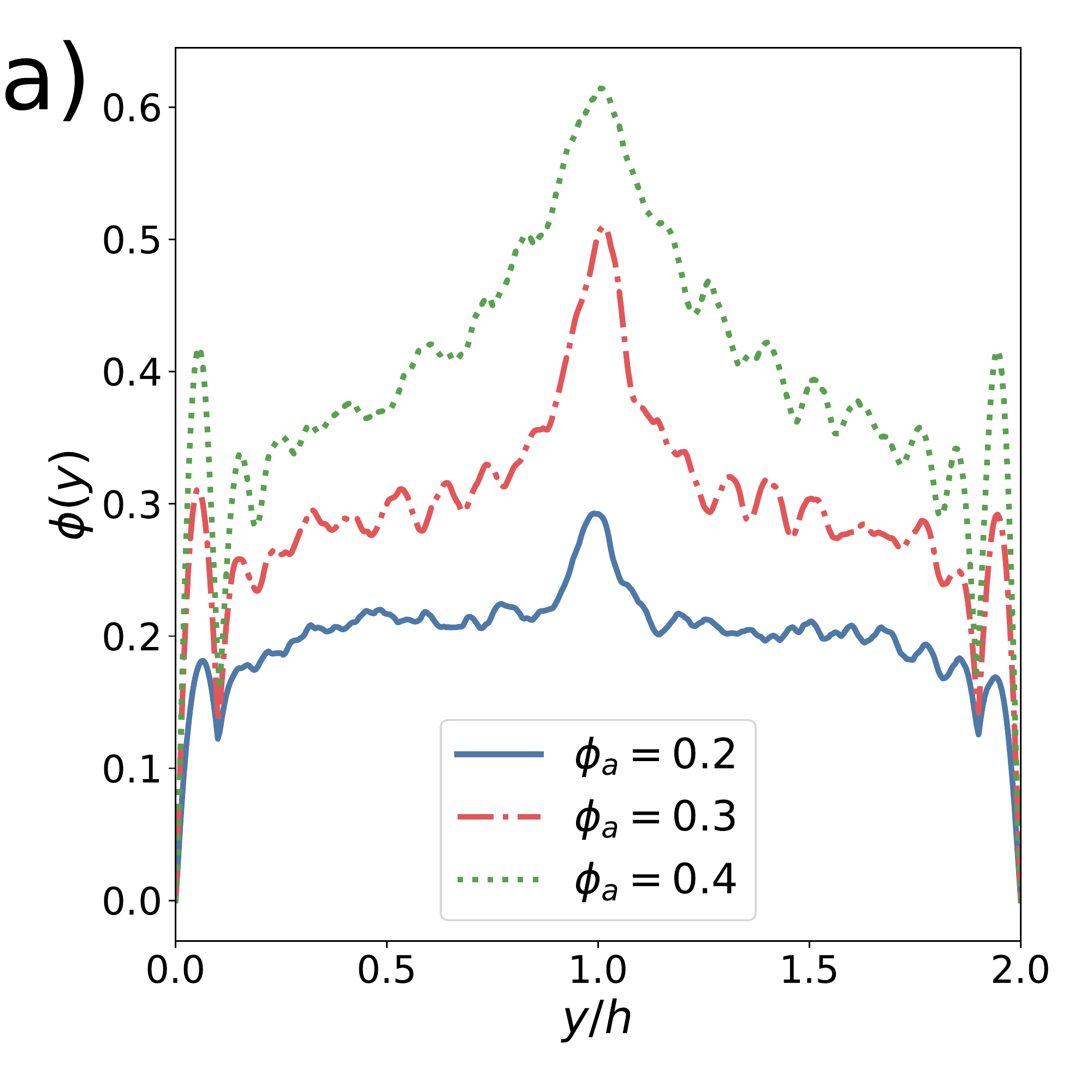}}
{\includegraphics[width=0.45\linewidth]{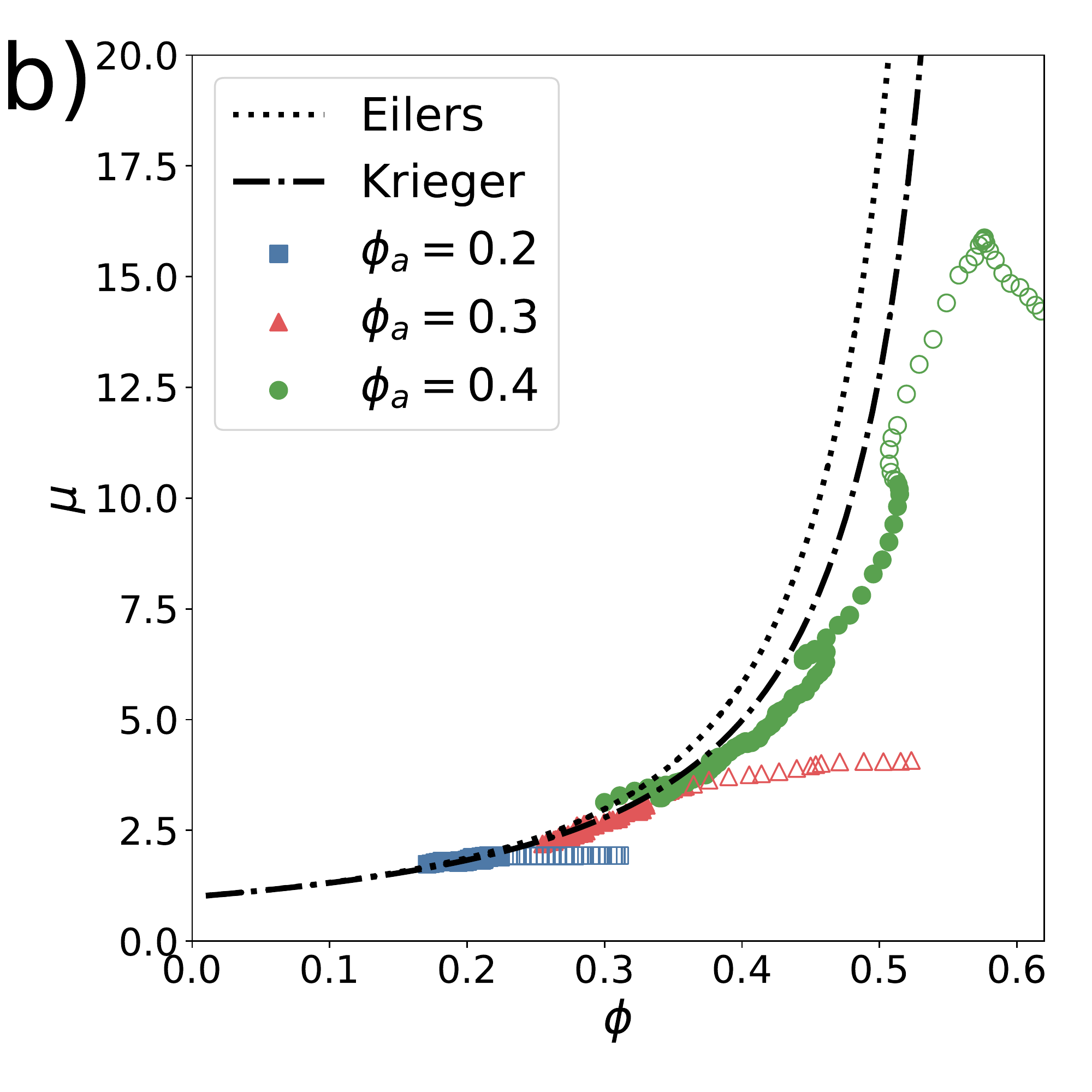}}
 \caption{a) Final suspension local volume fraction profiles $\phi(y)$ in steady-state. b) Suspension viscosities learned from the PINN model as a function of the local volume fraction. Results are compared with the Eilers fit \cite{Ferrini1979, Stickel2005} and the Krieger fit \cite{Krieger1959}.  Filled symbols represent points that occur in the range $0h \leq y \leq 0.85h$, and empty symbols are in the range $0.85h \leq y \leq h$, to denote the deviations that occur from the theoretical values in the densely packed core of the channel. }\label{fig:visc_fits}
\end{figure}

The velocity profiles for the suspension flows with $\phi_a=0.2$ and 0.4 are shown in Figs. \ref{fig:suspension}a and d, respectively, and the local volume fractions $\phi(y)$ are depicted in  \ref{fig:visc_fits}a.
A key feature of suspensions is irreversible shear-induced migration of particles to areas of low shear rate \cite{Leighton1987}. 
 Particles in a suspension subjected to a Poiseuille flow will migrate to the channel centerline, greatly increasing the volume fraction at the centerline until it reaches the maximum close-packing limit, as shown in Fig. \ref{fig:visc_fits}a. This migration also impacts the velocity profile, resulting in a flattened parabola shape that is observed in Figs. \ref{fig:suspension}a and d.

As in the analysis of polymer melts above,  we use the inverse PINN to find the viscosity $\mu(u_y(y))$ 
 by approximating $u$ and $\mu$ with $\hat{u}(y;\theta)$ and $\hat{\mu}(\hat{u}_y(y;\theta);\gamma)$ DNNs trained by minimizing the loss function  (\ref{loss_mu}) with $\omega_1 = \omega_2 =\omega_3 =1$ and $\omega_4=0$.
We use $N_u=401$ measurements of the velocity profile $u(y)$ from the FCM simulations. Because the velocity profiles from the simulations (see Fig. \ref{fig:suspension}a and d) deviate from the flattened-parabola shape near the walls due to particle layering, a phenomenon that cannot be described by  Eq.  (\ref{gov_eq}), we train the PINN with velocity data in the range $y \in [0.25y/h, 1.75y/h ]$, but still impose the zero Dirichlet BCs for $u$ at $y = 0, H$. 

 Figs. \ref{fig:suspension}a and  \ref{fig:suspension}d compare the velocity profiles of the suspension flow observed in the numerical simulations and are approximated with the $\hat{u}(y;\theta)$ DNN for $\phi_a=0.2$ and 0.4, respectively.  The $\hat{\mu}(\hat{u}_y(y;\theta);\gamma)$ DNN and the viscosity estimated from the numerical experiments are plotted in Figs. \ref{fig:suspension}b and  \ref{fig:suspension}e. 
 The viscosity $\mu(u_y)$ for the FCM simulations is found by computing  $u_y(y)$ and $\phi(y)$ from the simulation data, assuming that $\mu(u_y) = \eta_s(\phi(u_y)) \eta_f$ and using the Eilers formula
$\eta_s(\phi) = \left(1 + \frac{5\phi}{4\left(1-\frac{\phi}{\phi_c}\right)}\right)^2$  \cite{Ferrini1979, Stickel2005}. Here, $\eta_f$ is the fluid viscosity (which was set to unity in the FCM simulations) and $\phi_c$ is the maximum volume fraction of a suspension ($\phi_c = 0.62$ in the FCM simulations.)
We observe that the PINN method is able to accurately learn the velocity profile and captures the increase in viscosity at the channel centerline. Figs. \ref{fig:suspension}c and \ref{fig:suspension}f demonstrate that  the residuals are three orders smaller than $C = 0.0288$, indicating that the  $\hat{u}(y;\theta)$ and $\hat{\mu}(\hat{u}_y(y;\theta);\gamma)$ DNNs  satisfy Eq. (\ref{gov_eq}).   
 
Finally, we employ the inverse PINN method to evaluate $\eta_s$ as a function of $\phi$ and in Fig. \ref{fig:visc_fits}b compare it with the  Eilers model and the Krieger  model 
	$\eta_s(\phi) = \left(1-\frac{\phi}{\phi_c}\right)^{-2.5\phi_c}$ \cite{Krieger1959}.
In the PINN method, we compute $ \eta_s(\phi)$ using the $\hat{\mu}(\hat{u}_y(y;\theta);\gamma)$ and $\hat{u}(y;\theta)$ DNN models of viscosity and velocity and $\phi(y)$ observed in the FCM simulations. The considered empirical models predict similar $\mu$ values for $\phi<0.35$ away from the channel centerline. The empirical models assume that $\mu(\phi)$ is independent of $\phi_a$. Fig. \ref{fig:visc_fits}b shows that the PINN predicted $\mu(\phi)$ functions agree with the empirical models  for small $\phi$ for all considered $\phi_a$. For large $\phi$, the PINN estimated $\mu(\phi)$ relationships depend on $\phi_a$ and deviate from all considered empirical models. There are several reasons for the disagreement between the PINN estimated and empirical viscosity models.  At high volume fractions near the close-packing limit, $\phi_c$, the particle movements are highly correlated leading to non-locality of the particle forces. Therefore, Eq.  (\ref{gov_eq}) breaks down for dense suspensions at the centerline. Additionally, Eq. (\ref{gov_eq}) with a constitutive relationship of the Eilers or Krieger analytical forms predicts that the suspension will reach maximum packing with $\phi = \phi_c$ at the centerline  that is independent of the average volume fraction $\phi_a$ \cite{Guazzelli2018}. However, FCM simulations \cite{Yeo2011} and experiments \cite{Lyon1998} show that the volume at the centerline varies with the initial average volume fraction of the system.  

In conclusion, we have extended the PINN method for learning unknown physics, including the functional dependence of viscosity on the shear rate and other properties of fluids using indirect measurements such as fluid velocity and volume fraction. We have also demonstrated that once an accurate DNN approximation of the viscosity is available, the PINN method can be used to model non-Newtonian flow without any data except the boundary conditions.


\bibliography{mybib}
\end{document}